\documentclass[letterpaper,twocolumn,10pt]{article}
\usepackage{usenix}
\usepackage[available]{usenixbadges}
\usepackage{booktabs}
\usepackage[table]{xcolor}
\usepackage{graphicx}
\usepackage{siunitx}
\usepackage{subcaption}
\usepackage[most]{tcolorbox}
\usepackage[margin=1in]{geometry}
\usepackage{dcolumn}
\usepackage{listings}
\usepackage{enumitem} %USE FOR SUBMISSION
\usepackage{soul}
\usepackage{threeparttable}
\usepackage{xurl}
\usepackage{fvextra}
\DefineVerbatimEnvironment{verbatim}{Verbatim}{breaklines,breakanywhere}
\usepackage{hyperref}
\hypersetup{
    colorlinks=true,urlcolor=purple,linkcolor=cyan,citecolor=cyan
}

\lstdefinestyle{json}{
  basicstyle=\ttfamily\footnotesize,
  backgroundcolor=\color{gray!5},
  breaklines=true,
  frame=single
}

\newcolumntype{d}[1]{D{,}{,}{#1}}

\newif\ifdraft
  \drafttrue %% turn on comments
\ifdraft
    \newcommand{\acv}[1]{\textcolor{blue}{[[Alejandro: #1]]}}
    \newcommand{\nicolasc}[1]{\textcolor{brown}{[[Nicolas: #1]]}}
    \newcommand{\man}[1]{\textcolor{magenta}{[[Manoel: #1]]}}
    \newcommand{\todo}[1]{\textcolor{red}{\textbf{TODO: #1}}}
    
\else
    \newcommand{\acv}[1]{}
    \newcommand{\nicolasc}[1]{}
    \newcommand{\man}[1]{}
    \newcommand{\todo}[1]{}
    
\fi

\begin{document}
%-------------------------------------------------------------------------------

\date{}

\title{\Large \bf Chameleon Channels: Measuring YouTube Accounts\\Repurposed for Deception and Profit}

\author{
{\rm Alejandro Cuevas}\\
Carnegie Mellon University
\and
{\rm Manoel Horta Ribeiro}\\
Princeton University
\and
{\rm Nicolas Christin}\\
Carnegie Mellon University}

\maketitle

%-------------------------------------------------------------------------------
\begin{abstract}

Online content creators spend significant time and effort building their
user base through a long, often arduous process that requires finding
the right ``niche'' to cater to. So, what incentive is there for an established
content creator known for cat memes to completely reinvent their page channel
and start promoting cryptocurrency services or covering electoral news
events? And, if they do, do their existing subscribers not notice?

We explore this problem of
\textit{repurposed channels}, whereby a channel changes its
identity and contents. We first characterize a market for ``second-hand'' social media
accounts, which recorded sales exceeding USD~1M during our
6-month observation period. Observing YouTube channels (re)sold over these 6~months,
we find that a substantial number (53\%) are used to disseminate
policy-sensitive content, often without facing any penalty. 
Even more surprisingly, these channels seem to
gain rather than lose subscribers.

We estimate the prevalence of channel repurposing ``in the wild,'' using two snapshots of $\sim$1.4M YouTube
accounts sampled from an ecologically valid proxy. In a 3-month period, we estimate that $\sim$0.25\%
channels were repurposed. 
Through a set of experiments, we confirm that these repurposed
channels share several characteristics with sold channels---mainly,
the fact that they have a significantly high presence of policy-sensitive content. Across repurposed channels, we find channels similar to those used in influence operations, as well as channels used for financial scams. Repurposed channels have large audiences; across two observed samples, repurposed channels collectively held $\sim$193M and $\sim$44M subscribers. We reason that purchasing an existing audience and the credibility associated with an established account is advantageous to financially- and
ideologically-motivated adversaries. This phenomenon is not exclusive to
YouTube and we posit that the market for cultivating organic audiences
is set to grow, particularly if it remains unchallenged by mitigations,
technical or otherwise.

\end{abstract}

%-------------------------------------------------------------------------------

\section{Introduction}

In 2022, more than 22,000 users suddenly found themselves following @HonorColoradoOk---an account affiliated with Paraguay’s ruling political party---despite never opting in~\cite{abc_infovacunate_honorcolorado_2022}. Overnight, timelines were flooded with political content, just a year ahead of the presidential elections. The account had previously operated as @InfoVacunatePy, a government-linked source for vaccination site updates, but was allegedly sold and \textit{repurposed}. The abrupt change in identity, amplified by the polarized political landscape, quickly drew user complaints and media attention. Within days, the owners removed the account, but not before sparking public outcry and exposing how easily an established audience can be redirected toward new---sometimes questionable---content. Yet, in many cases, these identity changes go unnoticed.

The ability to repurpose accounts has long enabled the buying and selling of social media identities, historically conducted through private exchanges on forums and chat applications. These transactions were opaque and difficult to systematically study. In recent years, however, the rise of dedicated marketplaces combined with the reduced costs of content creation afforded by AI has dramatically expanded both the scale and visibility of this ecosystem. A recent study documented 38,000 accounts for sale, with an advertised value exceeding USD~64 million~\cite{beluri2024exploration}. Fameswap, one of the most active marketplaces, has now ninefold more active listings since its initial observation by Chu et al. in 2022~\cite{chu2022tube}.

\begin{figure*}[t!]
    \centering
    \begin{subfigure}[t]{0.561\textwidth}
        \includegraphics[width=\textwidth]{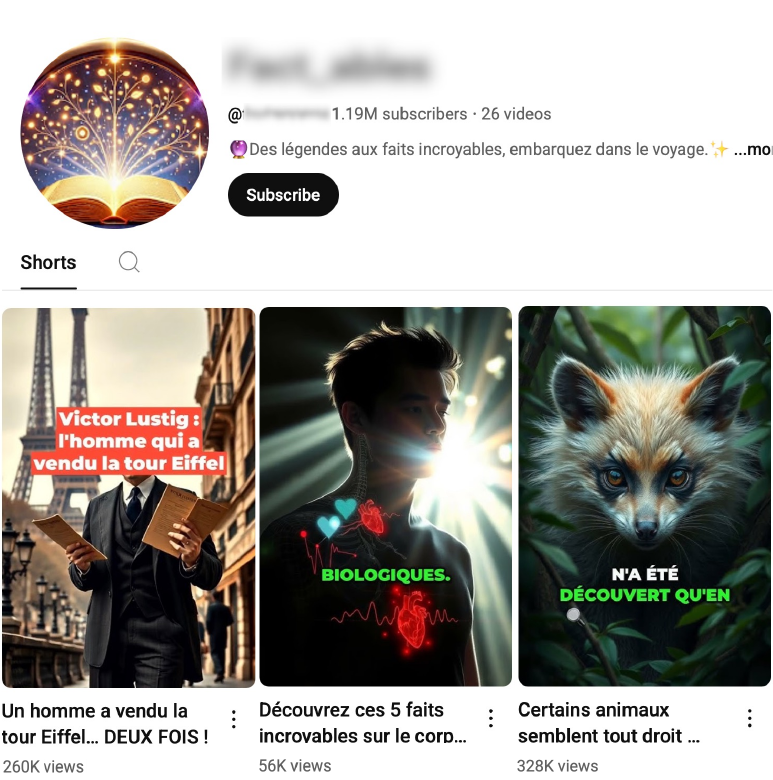}
        \caption{A channel when listed for sale.}
        \label{subfig:pre_reuse}
    \end{subfigure}
    \hfill
    \begin{subfigure}[t]{0.419\textwidth} % 4.48 / total
        \includegraphics[width=\textwidth]{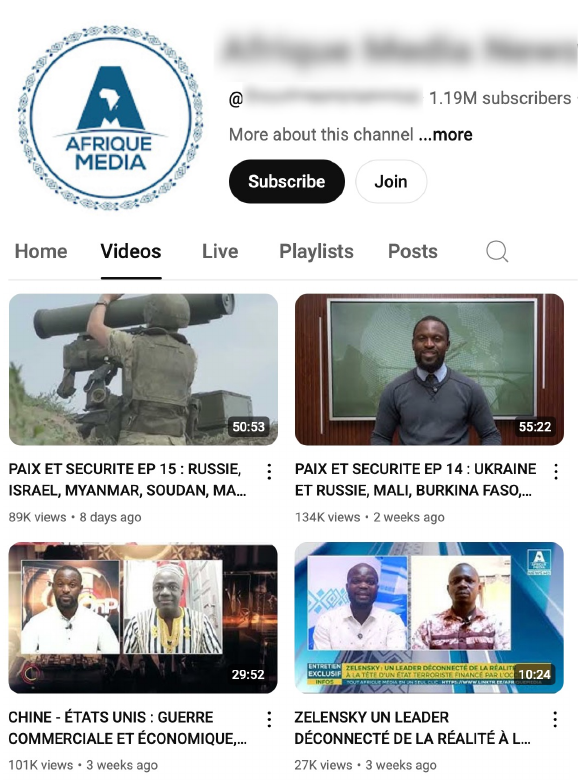}
        \caption{Channel after reuse.}
        \label{subfig:post_reuse}
    \end{subfigure}
    \caption{On the left, a channel listed for sale on Fameswap about entertaining facts with 1.19M subscribers. Videos are created predominantly with AI tools. This channel would later change to a news channel discussing political issues with no trace of its previous identity (on the right). Note, the channel on the left has been reconstructed using metadata. To not drive traffic to the channels, handles and titles are blurred.}
    \label{fig:2_part_real_example}
\end{figure*}

A key question is whether platforms like Fameswap constitute a new
paradigm for online engagement---as opposed to just a new packaging for
already known artificial engagement strategies~\cite{nevado-catalan2023analysis}. If these ready-made
accounts represent a new type of product, what are the implications for
the platforms whose accounts are commercialized? We address this question by observing how these accounts are used after being sold. We examine this question on YouTube by leveraging two measurement advantages. First, the rise of increasingly transparent account marketplaces provides direct visibility into accounts offered for sale. Second, YouTube assigns each channel a persistent identifier that remains fixed even when the channel changes its handle, title, description, or content, allowing us to track what we call \textit{repurposing}, illustrated in Figure~\ref{fig:2_part_real_example}.

Repurposing channels may alleviate a critical
constraint for financially and ideologically motivated adversaries:
rapid access to content distribution without the resource expenditure required
to cultivate an audience from scratch. We explore this possibility through three exploratory questions:

\begin{itemize}[itemsep=0pt, parsep=1pt, topsep=0.7pt, partopsep=0.7pt]
  \item \textbf{RQ1:} \textit{How are channels used after being repurposed on YouTube?}
  \item \textbf{RQ2:} \textit{How prevalent is repurposing on YouTube?}
  \item \textbf{RQ3:} \textit{What are indicators of repurposed channels?}
\end{itemize}

We then test the following hypothesis:

\begin{itemize}[topsep=0.7pt, partopsep=0.7pt]
    \item \textbf{H1}: \textit{Repurposed accounts (either advertised for sale or obtained in the Social Blade sample) disseminate policy-sensitive content at significantly higher rates than non-repurposed accounts.}
\end{itemize}

To investigate these questions, we collect YouTube accounts advertised for sale from Fameswap, one of the largest social media account marketplaces~\cite{beluri2024exploration}, and from Social Blade, an analytics platform that tracks 68M+ YouTube accounts (\S~\ref{sub:data-collection}). We then define and annotate channel repurposing and presence of policy-sensitive content (i.e., content which, due to its sensitive nature and potential for abuse, is regulated by YouTube's policies~\cite{youtube2024youtubes}) in \S\S~\ref{sec:channel_reuse}--\ref{sec:content_analysis}. In \S~\ref{sec:fameswap-results}, we explore the impact of repurposing on growth, suspension, and content changes through repurposed Fameswap accounts. Then, by labeling repurposing on the Social Blade sample, we estimate the prevalence of repurposing across YouTube in \S~\ref{subsec:prevalence}. Lastly, we compare repurposing indicators and differences in content by comparing repurposed channels (both from Fameswap and Social Blade) against a non-repurposed control group (\S~\ref{sec:indicators_of_reuse}).

We find that repurposed accounts continue to accrue subscribers after repurposing, that most of them are not suspended, and that repurposing is prevalent. We estimate that 0.25\% of all accounts indexed by Social Blade were repurposed between January and March 2025 and argue that this estimate generalizes to YouTube as a whole. We find that repurposed channels post higher rates of policy-sensitive content, especially ideological and financial content. During our study, we observed that repurposed Fameswap accounts had 193M+ subscribers and repurposed Social Blade accounts had 43M+ subscribers. Purchasing and repurposing accounts seems to be a viable way to quickly distribute content which may affect the integrity of online platforms, particularly if the content may result in financial harm to users or is part of a political influence operation. We argue that these findings generalize to other platforms where accounts are traded on secondary markets and where modifying profile indicators (e.g., handles or usernames) is trivial and difficult to detect.

\section{Related Work}
\label{sec:related} 
There is considerable related work in areas adjacent to this study,
which we will discuss next. In particular, we explain how our study differs
from these previous efforts.

\noindent\textbf{Abusing Reputation Systems.} Online reputation refers to metrics that signal trustworthiness. Reputation systems---reviews, ratings, etc.---were first developed in online marketplaces and instrumental to their success~\cite{tadelis2016reputation}. These systems were later adapted by social media to incentivize various behaviors~\cite{seering2020reconsidering,kraut2012building,anderson_steering_2013} and signal authenticity~\cite{vaidya_does_2019}. Today, they take the form of likes, followers, badges, and similar indicators. These indicators are valuable targets for manipulation. Researchers have documented widespread abuse across platforms~\cite{mayzlin_promotional_2014,luca_fake_2016,xu_e-commerce_2015,stringhini2013follow,nevado-catalan2023analysis}. Misuse has evolved from bots and compromised accounts~\cite{stringhini2012poultry,thomas2013trafficking,egele2013compa,cao_aiding_2012} to more human-like tactics: crowdsourced engagement~\cite{wang2012serf,yao_automated_2017,motoyama2011dirty}, automation of real user accounts~\cite{dekoven2018following}, collusion rings~\cite{weerasinghe2020pod,farooqi2017measuring}, and click farms~\cite{grohmann2022click,drott2020fake}.

Account hijacking has similarly shifted toward high-value targets. Rather than compromising many small accounts, actors now focus on fewer but influential ones~\cite{egele2017towards}. This shift may be tied to the ease of monetizing influence, particularly in finance and crypto spaces. Notable examples include the SEC’s hijacked account used for market-moving tweets~\cite{justice2025alabama} and Vitalik Buterin’s account used to spread phishing links~\cite{knight2023hack}.

Reputation abuse also occurs without hijacking. Liu et al.~\cite{liu2024give} show that scammers use legitimate-looking YouTube channels (e.g., ``Ripple XRP'' with 114,000 subscribers) to host ephemeral livestreams and direct viewers to off-site scams.

\noindent\textbf{Licit Yet Harmful Online Influencers.} Influential accounts that
promote potentially harmful content but within gray areas have become
an increasingly prevalent issue. For example, many reported cases of
influencers promoting ``meme coins'' have left viewers with little recourse
after the coin collapses and they incur losses~\cite{clark2024hawk,khalili2025memecoin}. It is unclear whether YouTube’s terms of service prohibit influencers from promoting speculative financial instruments such as meme coins, in part because the legality of such promotions remains unsettled. Nonetheless, their financial impact on viewers is clear.
Similarly, influential accounts have also been used to disseminate
problematic narratives, such as smear campaigns, during election periods. The latest case occurred just last year, when a criminal investigation
linked payments from a Russian disinformation campaign to American
influencers~\cite{grevygotfredsen2024tenet}. This example also
illustrates the increasing interplay between financially motivated
actors (in this case, influencers) and ideologically
motivated actors (state sponsors).

\begin{figure*}[h!]
    \centering
    \begin{subfigure}[t]{0.54\textwidth}
        \includegraphics[width=\textwidth]{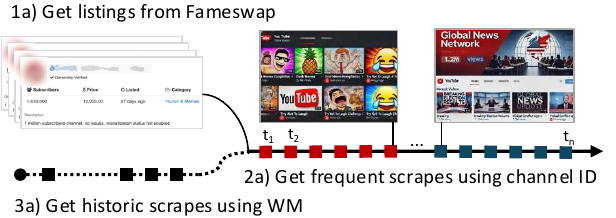}
        \caption{Collection procedure for channels listed on Fameswap.}
        \label{subfig:fswap_collection}
    \end{subfigure}
    \hfill
    \begin{subfigure}[t]{0.445\textwidth}
        \includegraphics[width=\textwidth]{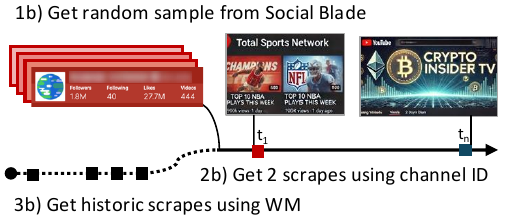}
        \caption{Collection procedure for reference sample.}
        \label{subfig:sb_collection}
    \end{subfigure}
        \caption{Data collection procedures. We collected Fameswap listings (1a) daily and a large sample of Social Blade channels (1b). We then scrape snapshots through time. On Fameswap, we scraped a channel every 3 days on average (2a). For Social Blade channels, due to the sample size, we collected a snapshot in January and later in March 2025 (2b).
        Finally, we obtain historical snapshots using the Wayback Machine (WM; see 3a and 3b).}
    \label{fig:collection}
\end{figure*}

\noindent\textbf{Disinformation and Underground Markets.}
Services to assist or implement disinformation campaigns have
often overlapped with underground markets and
communities~\cite{wallis2021influence,grohmann2024disinformation,madung2021kenya,guardian2023disinformation}. 
On one hand, these marketplaces
specialize in the dissemination of content that is frequently at odds
with platform guidelines, for example, advertising restricted products
(e.g., pharmaceuticals, tobacco products, unregulated gambling websites,
etc.), including disinformation. On the other hand, these marketplaces
also provide tools and infrastructure that can aid disinformation
operators (e.g., coordinated engagement, bulletproof hosting, DDoS
protection, advertising networks, etc.)~\cite{han2022infrastructure}.
Most importantly, off-platform sources of monetization are
an important driver for channels with problematic content to
thrive~\cite{chu2022tube,hua2022characterizing,ballard2022conspiracy}.

\noindent\textbf{Domain and Handle Reuse.} Our work shares many
similarities with work on residual trust across domains and
namesquatting~\cite{so2022domains,lever2016domain}. Most
related to our work, Elmas et al. study account misleading repurposing on Twitter, whereby an account changes profile attributes to use the account for a new purpose while retaining its followers~\cite{Elmas_Overdorf_Aberer_2023}. They propose a supervised learning method for detection and find common behavioral patterns across repurposed accounts (e.g., repurposing after inactivity or tweet deletion~\cite{Elmas_Overdorf_Aberer_2023}. Our work builds upon theirs but extends it by estimating the extent to which repurposing occurs across a whole platform and comprehensively quantifying the types of content that are posted after repurposing. Also related is work by Mariconti et al.\ who study the reuse of profile names on
on Twitter~\cite{mariconti2017whats}. Their study focuses on
accounts that register handles that had been de-registered to
capture the residual reputation and backlinks pointing to those
handles~\cite{mariconti2017whats}. In our case, repurposed channels carry all previous followers.

\section{Background}

Fameswap is an online marketplace for buying and selling accounts on YouTube, TikTok, Instagram, Twitter/X, and websites. Recent measurements identify it as the largest such market by number of sellers~\cite{beluri2024exploration}. The site resembles a standard e-commerce platform: each account has a dedicated listing with a description, price, seller reviews, and a content category (e.g., humor, sports). Buyers can bid, and the platform offers escrow, dispute resolution, and account verification—features similar to marketplaces like eBay. Fameswap earns revenue from escrow fees (3\% or a USD~50 minimum) and premium accounts, which offer advanced search, additional metrics, lower fees, and more. Payments are accepted via PayPal, wire transfers, and cryptocurrencies (e.g., BTC, ETH, USDC). Fameswap verifies
account ownership by providing sellers with a unique randomized string that
should be placed in the advertised social media profile. For example, a
seller receives the string \texttt{aSgc5H3s} and places it temporarily in their
YouTube channel description, allowing Fameswap to verify the ownership, similar to using TXT records to verify domains.

Fameswap listings link the corresponding channel's YouTube ID. Different from channel
titles, handles (``\texttt{@SomeChannel}'') are unique and can be accessed
through ``\nolinkurl{www.youtube.com/@SomeChannel}~\cite{youtube_handles_2022}.'' Handles resolve to a
unique channel ID, allowing a channel to change its handle but
keep pointing to the same channel. YouTube's channel ID  is a string that begins with ``UC'',
followed by 22 characters (letters, numbers, dashes, and underscores).
Channel IDs can also be used to access a channel in the form:
``\nolinkurl{www.youtube.com/channel/UC...}''. While an account can
change its handle, its channel ID cannot change, which allows us
to monitor accounts over time.

The market for ready-made accounts (as opposed to the market for artificial engagement~\cite{nevado-catalan2023analysis})
has remained largely understudied, only receiving a brief overview by
Chu et al.~\cite{chu2022tube} in their study of illicit alternative monetization strategies on YouTube. In their study, they identified five marketplaces for social media accounts: SWAPD\cite{swapd2025}, Accs-Market.com~\cite{accsmarket2025}, Trustiu (now inactive), ViralAccounts~\cite{viralaccounts2025}, and Fameswap~\cite{fameswap2025}. Beluri et al., expand on Chu et al.'s work by conducting a more comprehensive financial characterization of social media account markets\footnote{Their study became public after our data collection began and is referenced retrospectively.}\cite{beluri2024exploration}. Their results indicate that Fameswap is a leading marketplace and thus an appropriate candidate for study~\cite{beluri2024exploration}. In addition to the platforms described in these studies, we note that a substantial amount of trade occurs in other platforms that are not described, such as the forum OGUser (similar to SWAPD) and Telegram channels.

\section{Methods}
\label{sec:methodology}

\begin{figure*}[!ht]
    \centering
    \includegraphics[width=\linewidth]{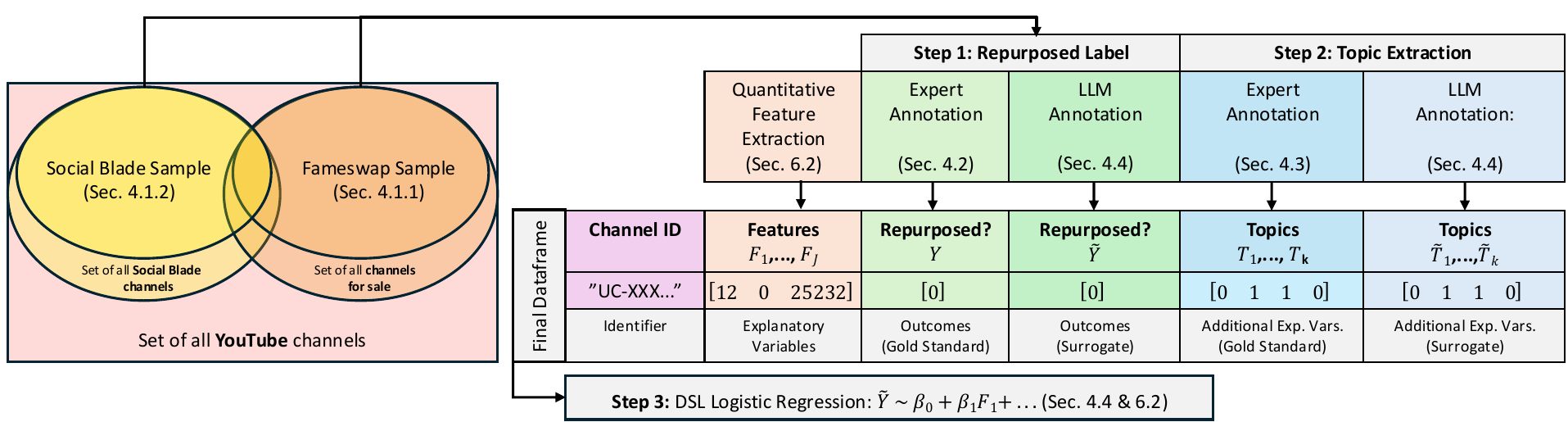}
    \caption{End-to-end experimental procedure, including samples, annotations, featurization, and regression.}
    \label{fig:full-method}
\end{figure*}

We measure the repurposing of YouTube accounts and their participation in
disseminating policy-sensitive content through the lens of sold accounts
extracted from a social media marketplace (Fameswap) and by observing
handle changes and channel repurposing in the wild from an ecologically valid proxy (Social Blade). We
structure our study in three parts, as depicted in Figure~\ref{fig:full-method}. First, we define and characterize channel repurposing across Fameswap channels (\S~\ref{sec:channel_reuse}).
Second, we develop a codebook for topic extraction and use it to annotate repurposed Fameswap channels and observe content dissemination after repurposing (\S~\ref{sec:content_analysis}). Using these annotations in \S~\ref{sec:fameswap-results}, we characterize the Fameswap marketplace and the repurposed accounts obtained from there. Third, we observe repurposing in the wild and estimate its prevalence. We then use a framework for deriving statistically valid estimates using textual annotations and quantitatively test what features are indicative of repurposing across samples (\S~\ref{sec:gen-and-prev}).

\subsection{Data Collection and Sources}
\label{sub:data-collection}
To carry out all our experiments, we collected, labeled, and categorized YouTube metadata from two sources, Fameswap and Social Blade, as shown in Figure~\ref{fig:collection}.
We provide a summarized description of our samples, as well as instructions for data access and replication in Appendix~\ref{sec:open_science}. An additional investigation using the Wayback Machine is discussed in Appendix~\ref{apx:wayback}. Lastly, ethical considerations are discussed in Appendix~\ref{sec:ethics}.

\subsubsection{YouTube Channels For Sale on Fameswap}
\label{subsubsec:fswap_channels} 
As illustrated in Figure~\ref{subfig:fswap_collection}, we conducted daily scrapes
of Fameswap from October 26, 2024, to March 31, 2025. The Fameswap
interface provides a paginated list of all account listings. We scraped
all historical listings. This set includes all listings not deleted
or hidden from the website. Using the YouTube IDs linked to each listing, we conducted regular scrapes of all collected channels. We scraped channels every 3 days
on average. As of March 31, 2025, we collected 4,641 YouTube channel
IDs advertised for sale, each with about eight observations per month.
Form Fameswap listing, we scraped their YouTube
channel using the YouTube Data API v3.

\subsubsection{Collecting a Large Ecologically Valid Sample of YouTube through Social Blade}
\label{subsubsec:socialblade_channels}
To estimate the prevalence of channel repurposing and identify repurposing indicators beyond marketplaces, we aimed to observe this phenomenon ``in the wild.'' Randomly sampling YouTube is difficult due to the lack of a centralized channel directory and method to enumerate channel IDs~\cite{hortaribeiro2021youniverse}. Horta Ribeiro and West address this by sampling from Social Blade~\cite{hortaribeiro2021evolution}, an analytics site that has indexed over 68M channels in the past 17 years~\cite{socialblade_website}. Social Blade is a widely used reference for creators~\cite{socialblade_wikipedia} and has informed multiple YouTube studies~\cite{ribeiro2020auditing,freire2022understanding}. While Social Blade's crawling process is not public, it likely favors more popular channels. This, however, is acceptable given that our focus is content creators with larger audiences.

To conduct this study, we obtained a dataset of YouTube IDs collected by Horta Ribeiro and West. We randomly sampled 1.4M channels and conducted an initial scrape from December 23, 2024, to January 21, 2025 (Figure~\ref{subfig:sb_collection}). We collected metadata from 1,397,586 channels and 139M corresponding videos. A second scrape, from March 21–31, 2025, captured updated metadata and 20M new videos. Only 1,351,912 channels returned data (a 3.3\% deletion rate). We used \texttt{yt-dlp}, a common tool for archiving YouTube content~\cite{youtube_dl_wikipedia,reddit_youtubedl,retkowski-waibel-2024-text}. While \texttt{yt-dlp} is commonly used in academic projects, we discuss potential ethical implications in Appendix~\ref{sec:ethics}.

\subsection{Labeling Repurposed Channels}
\label{sec:channel_reuse}

Our first goal is to detect channels that have been repurposed, such as the example shown in Figure~\ref{fig:2_part_real_example}. In this figure, we observe how a channel that was posting primarily entertainment facts then becomes a channel that shares political news. Repurposing a channel allows an individual to portray herself as legitimate,  distribute content to an existing
audience, or potentially increase the chances that the
recommendation algorithm amplifies new content~\cite{yoganarasimhan2012impact,ling2022slapping}. We first apply a qualitative approach to define what constitutes a channel being repurposed. We use these qualitative insights
to develop an LLM prompt which we can then use to scale our annotations.

Determining whether a channel has been repurposed is an entity matching problem. We can say that a tuple representing a paired observation in time of a given channel constitutes a channel being repurposed if both observations do not refer to the same identity. However, given that we do not have ground truth for channel identities, we first create a definition based on human perception and then enhance this procedure using an LLM. Because LLMs have been trained on large text corpora, LLMs are well-suited for entity matching problems because they can identify
connections between entities a human annotator may miss~\cite{peeters2023using}.

To create a definition, we followed a three-stage process. First, the lead author manually monitored a random sample of 104 channels (10\%) that changed handles after being listed for sale. Each channel was visited weekly over a month to observe changes in content, metadata, and activity. We created standardized weekly snapshots capturing each channel’s videos, thumbnails, titles, timestamps, and descriptions (Appendix~\ref{apx:sup_materials}). Second, using open coding, two researchers independently analyzed a new random sample of 54 channels (5\%) using these documents. Each annotator coded the channels following the prompt: ``was the channel repurposed?'' Coders discussed their annotations to identify edge cases and refine definitions. Disagreements arose around dormant channels, subtle content shifts, and stylistic similarities with different themes (e.g., channels that had same style of content but around different themes). This coding was exploratory, thus we did not compute agreement scores. Coders prioritized changes in handles and titles, using descriptions and video content as secondary evidence. Minor changes (e.g., \texttt{@HealthReporter} to \texttt{@TheHealthReporter}) were not considered repurposed. Significant discrepancies (e.g., \texttt{@HealthReporter} to \texttt{@247News}) prompted checks for overlap in topics, URLs, or referenced entities. If no continuity was found, the channel was flagged as repurposed. A substantial shift in video content (e.g., religious to crypto content in another language) served as a confirmatory signal but was not sufficient on its own. In the third step, the lead author and an external coder with a large social media following and understanding of the content creation ecosystem reviewed another 54 channels (5\%) as a sanity check.

\tcbset{
  mybox/.style={
    colback=gray!10,
    colframe=black,
    arc=4mm,
    boxrule=0.5pt,
    width=\columnwidth,
    sharp corners=south, % optional for visual style
    fonttitle=\small\bfseries,
  }
}

\begin{tcolorbox}[mybox, title=Definition: Repurposed Channel]
We consider a channel to have been repurposed from time $t$ to $t+1$ (two
observations in time) if there is no perceivable association between
the channel's new handle and new title from its prior handle and title.
Further, the channel description must not contain any references to its
previous identity (e.g., ``this channel was previously named X''), nor
any overlapping text, nor any pointers to the same resources (e.g., a
URL pointing to a social media account mentioned at time $t$).
\end{tcolorbox}

\subsection{Labeling Policy-Sensitive Content}
\label{sec:content_analysis}

Our second goal is to characterize the uses and understand motivations by observing the
content disseminated after repurposing a channel. We define as \textit{policy-sensitive}, content that due to their sensitive nature (e.g., financial, health, political content, etc.) and potential for abuse and harm, is regulated by YouTube's community guidelines and moderation practices~\cite{youtube2024youtubes,YouTubeMonetizationPolicies2025}. Informed by prior work\footnote{Scholars studying harmful content strongly advise researchers ``not to accept how platforms frame and define issues''~\cite{vidgen2019challenges}. This is because content guidelines are often broad and cover a wide range of topics, and definitions often lack
specificity~\cite{vidgen2019challenges}.} and our initial rounds of annotation, we focus on two main broad types of policy-sensitive content: financial and ideological. Additionally, also due to their prevalence in initial annotations, we labeled channels that contained AI-generated content, copyright infringement, content oriented towards children, and the presence of alternative forms of monetization (all of which are also governed by YouTube's policies, and hence policy-sensitive).

\subsubsection{Annotation for Codebook Development}

To develop our category annotation codebook, we adopted an empirical approach. We selected a set of restricted themes from YouTube's Community Guidelines and refined them in various coding rounds into narrower operational definitions. We discuss each of the content categories and then describe how we used an LLM to scale our annotations. Lastly, we explore the categories of content posted before and after repurposing.

We conducted three rounds of content annotation. First, the lead author selected 10\% of all repurposed channels ($n$=102) and created an initial codebook identifying which YouTube policies applied to the videos (based on metadata) contained in each channel. Using this initial codebook, the lead author and second author independently coded 54 repurposed Fameswap channels (5\% of reused
channels). Both authors have extensive experience with the YouTube
ecosystem.
Each researcher received a summary text document containing
all of the channel's observation snapshots. We provide an example of this
document in Appendix~\ref{apx:sup_materials}. After the first round of coding, the coders discussed the procedure and tweaked the codebook
to address the shortcomings. We also extracted additional codes from the
open-ended field if the content was governed by community guidelines and not yet captured by existing codes. 

The lead author and a member external to the research team with extensive knowledge about the creation of social media content did another round of coding. They coded
a new randomly sampled but non-overlapping 5\% of the dataset. The
goal was to identify and reduce ambiguity in the existing definitions. Together with this external member, we derived our final codebook. The final codebook is found in Appendix~\ref{apx:coding_guide}. Note that we did not compute agreement during these steps, since the goal of these steps was to iteratively create a codebook, not yet to use the codes for downstream analyses. We address the statisical soundness of our annotations in \S~\ref{subsub:annotation}. A discussion of ethics, as related to the annotation procedure, is found in Appendix~\ref{sec:ethics}.

\subsubsection{Content Categories}

\textbf{Ideological Content.} Based on YouTube's misinformation guidelines~\cite{youtube2025misinformation,youtube2025medical,youtube2025elections}, we created three
codes covering \textit{political}, \textit{medical}, and \textit{news}
content. News and political content often overlap (i.e., politcal news), but not always. We do not attempt to verify whether videos' content constitute misinformation or disinformation---the latter defined by its intentionality. We do not verify the veracity of claims, as we discuss in \S~\ref{sec:limitations}. However, the dissemination of political or news-related information through social media channels not associated to established news outlets is a hallmark of influence operations, regardless of the claims' veracity~\cite{woolley2023manufacturing}.

In addition to political and news content,
we noticed a substantial number of \textit{religious} videos in our sample. Although
religion is not a topic captured in community guidelines, we opted to
include it as a code given its frequent co-occurrence with geopolitical
content and frequently observed content involving religious leaders with
political appointments. Past work has also found religious videos to frequently co-occur with extremist and hateful content~\cite{albadi2018are,albadi2022deradicalizing}. Lastly, we included codes for \textit{extremist} content and \textit{manosphere} content, based on past work on radicalization pathways on YouTube~\cite{hortaribeiro2021evolution}.

\noindent\textbf{Financial Content.} On the financial side, we created a code for \textit{Gambling} content,
which is directly regulated by the guidelines~\cite{youtube2024illegal}. However, we note a
substantial amount of content related to making money online through a
variety of ways: pay-per-click websites, dropshipping, trading stocks,
cryptocurrencies, get-rich-quick schemes, and ironically, courses to
grow and monetize social media accounts. These topics are only partially covered by YouTube's guidelines (e.g.,
get-rich-quick schemes, fraud, broadly ``spam'', and broadly ``scams'')~\cite{youtube2024spam}.
Cryptocurrency-related content
was particularly prevalent. Informed by research on cryptocurrency scams and their potential to harm
users financially~\cite{kawai2023is,liu2024give}, we create a \textit{cryptocurrency} code. We
capture the rest of the aforementioned content under \textit{money-making
content}.

\noindent\textbf{Other Policy-Sensitive Content.} We also identified other types of content in our annotation that are covered by YouTube's guidelines, such as \textit{content that may infringe
copyright}~\cite{youtube2025copyright} and \textit{AI-generated content}~\cite{youtube_ai_2025}. AI-generated content is of particular importance given the raise of ``faceless YouTube channels'' and their adoption by scammers and spammers~\cite{diRestaGoldstein2024}. Lastly, we also identified content that appears to be geared towards children (e.g., content involved early childhood cartoons, nursery rhymes, and didactic activities like coloring) and pre-teens (e.g., content involved Roblox, Minecraft, and young influencers). We also labeled this content given YouTube's policies for content targeting minors and child safety~\cite{youtube2024child}. Lastly, we wanted to explore whether repurposed channels had a higher rate of alternative monetization strategies. Past work has found that alternative monetization strategies help channels dedicated to problematic content stay afloat ~\cite{hua2022characterizing}. Furthermore, we hypothesized that some forms of alternative monetization strategies, commonly discussed in underground forums~\cite{chu2022tube}, may be more associated with repurposing practices.

\subsection{Scaling Annotations with LLMs}
\label{subsub:annotation}

We use LLMs to scale our repurposing and content annotations for two reasons: (1) estimating repurposing across all Social Blade channels and (2) testing for content differences across samples. In the former case, we can use false positives/negatives to directly correct our estimate. For the latter, we employ a regression model. However, directly using LLM-generated labels in downstream regressions can introduce measurement error—mismatch between predicted and true labels~\cite{egami2023using}. Information on how to access the prompts is discussed in Appendix~\ref{apx:sup_materials}.

To correct for this measurement error, Egami et al. introduce a framework for statistically valid inference with
LLMs: Design-based Supervised
Learning (DSL)~\cite{egami2023using}. The process involves: (1) predicting labels with an LLM, (2) sampling a subset for expert annotation, and (3)
combining predictions and gold-standard labels in DSL regression.
This method requires that expert-coded samples have known, non-zero
sampling probabilities. This approach allows us to use mall-scale qualitative annotations with large-scale
LLM predictions, using the latter as ``surrogate'' classifiers.

We annotated our data using \texttt{gpt-4o-2024-11-20} (state-of-the-art at the time of writing) with temperature zero and top $p$ of one. We created LLM prompts based on our final codebooks and definitions following best practices suggested by Ziems et al.~\cite{ziems2024can}. Although there is still a lack of consensus on the optimal parameter choice for LLMs, recent work recognizes these parameters as the current standard~\cite{renze2024effect}. We validate our labels by comparing against expert annotations (also called ``gold-standard labels'') and describe the framework we employ to obtain statistically valid estimates using these annotations.

\noindent\textbf{Repurposing Labels.}
Using a prompt based on our repurposing definition, we annotated the Fameswap and Social Blade channels that we collected. We create a gold-standard dataset composed of a random 10\% subsample of human annotated channels for each sample. In our Social Blade sample, this annotation allows us to directly estimate the total set of repurposed channels between January and March 2025, using the false positive/negatives rates to derive a 95\% confidence interval. We present the results of this estimation in \S~\ref{subsec:prevalence}.

\begin{figure*}[t]
    \centering
    \includegraphics[width=\linewidth]{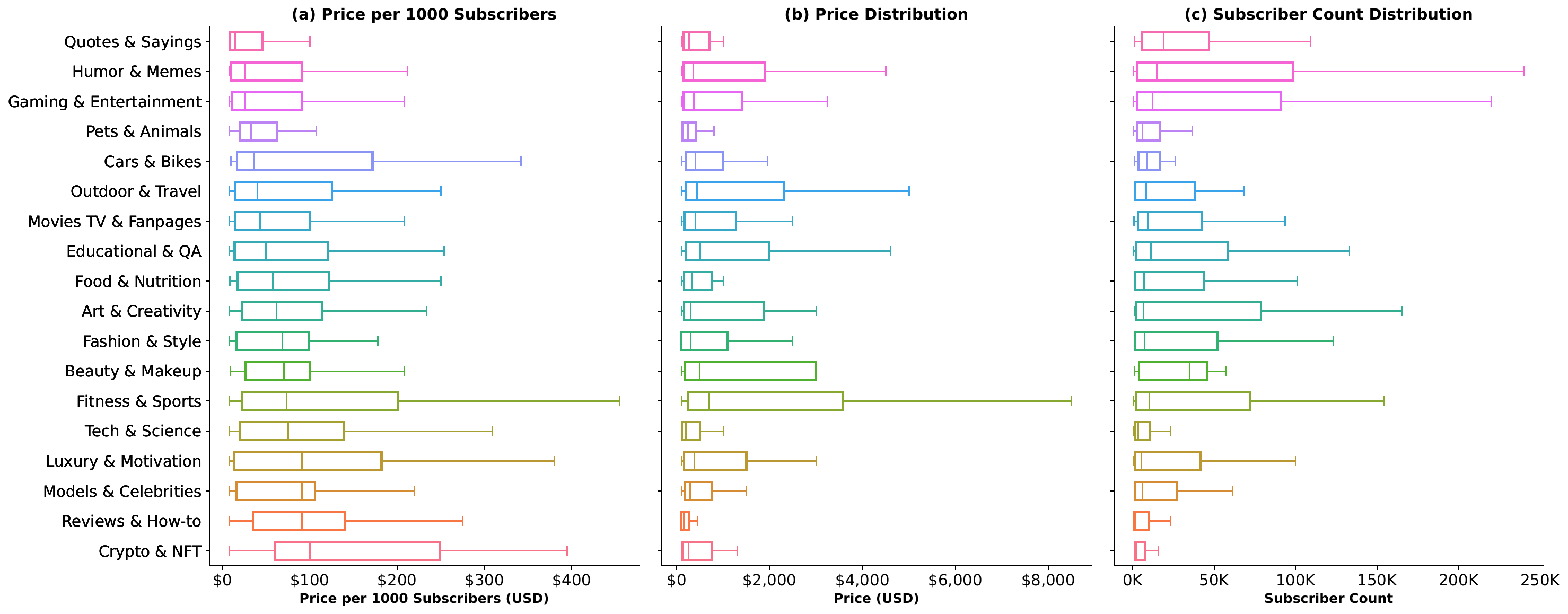}
    \caption{Distribution of listing prices, subscriber counts, and price per 1,000 subscribers for Fameswap listings. Outliers (Q3 + 1.5X IQR) are hidden. Categories are self-reported at the time of listing. Categories are ordered by median price per 1,000 subscribers.}
    \label{fig:fswap_distributions}
\end{figure*}

\noindent\textbf{Content Labels.} Similarly, using a prompt based on the annotation codebook we developed, we annotated three samples: repurposed Fameswap channels, repurposed Social Blade channels, and a random sample of non-repurposed Social Blade channels which we will use as a control group. For each channel, we provide the LLM with up to 50K tokens per channel, combining the channel description with video titles and descriptions. Most channels fit this limit. If a channel exceeds it, we (i) de-duplicate strings with $>$ 0.9 edit similarity to capture more diverse content, then (ii) randomly sample the remaining text until the total is $\leq$50K tokens. We do this to capture
a greater diversity of information, given that some channels post
many videos with the same or similar titles. We cap input at 50K (well below GPT-4o's 128 K window) because longer contexts seem to degrade classification accuracy~\cite{li2024long-context}.

For each of these samples, we annotate the content of a random 5\% ($n$=260, out of $n$=5,098) subsample drawn from the three samples. These channels had a combined total of 14,354 videos, which the lead author manually annotated. The annotations to the content are binary, indicating the presence of content.

\section{Fameswap Results}
\label{sec:fameswap-results}

We found 25,404 listings for social media accounts across all platforms, advertised worth a total of USD~366.7M and claiming over 2.6B followers. Between 2019-2021, Chu et al. found 3,112 listings on Fameswap~\cite{chu2022tube}, suggesting substantial growth to the present day. For YouTube specifically, we found 4,641 listings with 823M subscribers (confirmed via the YouTube API) and a combined listed price of USD~160.4M. 

To contextualize listing prices, we computed the cost per 1,000 subscribers---a common metric in engagement forums~\cite{nevado-catalan2023analysis}. As shown in Figure~\ref{fig:fswap_distributions}, this varies widely by category. The median cost on Fameswap in some categories is substantially higher than USD~16.52 per 1,000 subscribers found by Nevado-Catal\'an et al.~\cite{nevado-catalan2023analysis}, suggesting that other features like niche drive the valuation and that the audience for sale is pricier than a comparable sized inorganic audience.

Verifying sales remains difficult, as with other scraping-based estimates~\cite{cuevas-usenix-22,gibbon2024measuring}. We scraped 16,110 seller profiles, 7,000 more than Beluri et al.---like them, we also observed inflated pricing for some listings: 41 listings exceeded USD~100K, and 16 exceeded USD~1M~\cite{beluri2024exploration}. We identified 2,930 escrow transactions and 684 reviews. Only 1,590 transactions disclosed sale amounts, totaling USD~1.16M. Though modest, this likely represents only a fraction of actual sales, which also occur via forums, Discord, Telegram, and private deals on Fameswap~\footnote{We suspect public sales skew toward smaller, cheaper channels, as the average sale price (USD~731) is far below the average listing price (USD~5,400).}.

\subsection{Before and After Repurposing}
Using our repurposing definition (\S~\ref{sec:channel_reuse}) and our annotations (\S~\ref{subsub:annotation}), we labeled the set of Fameswap channels we collected. We identified 1,024 Fameswap channels (23\%) that were repurposed during our observation period. In total, these channels had a combined audience of 193M+ subscribers. In the following, we explore how long it took for these channels to change, the impact of repurposing on their existing audience, whether they were suspended, and the content to which they shifted.

\begin{figure}[t]
    \centering
    \includegraphics[width=\linewidth]{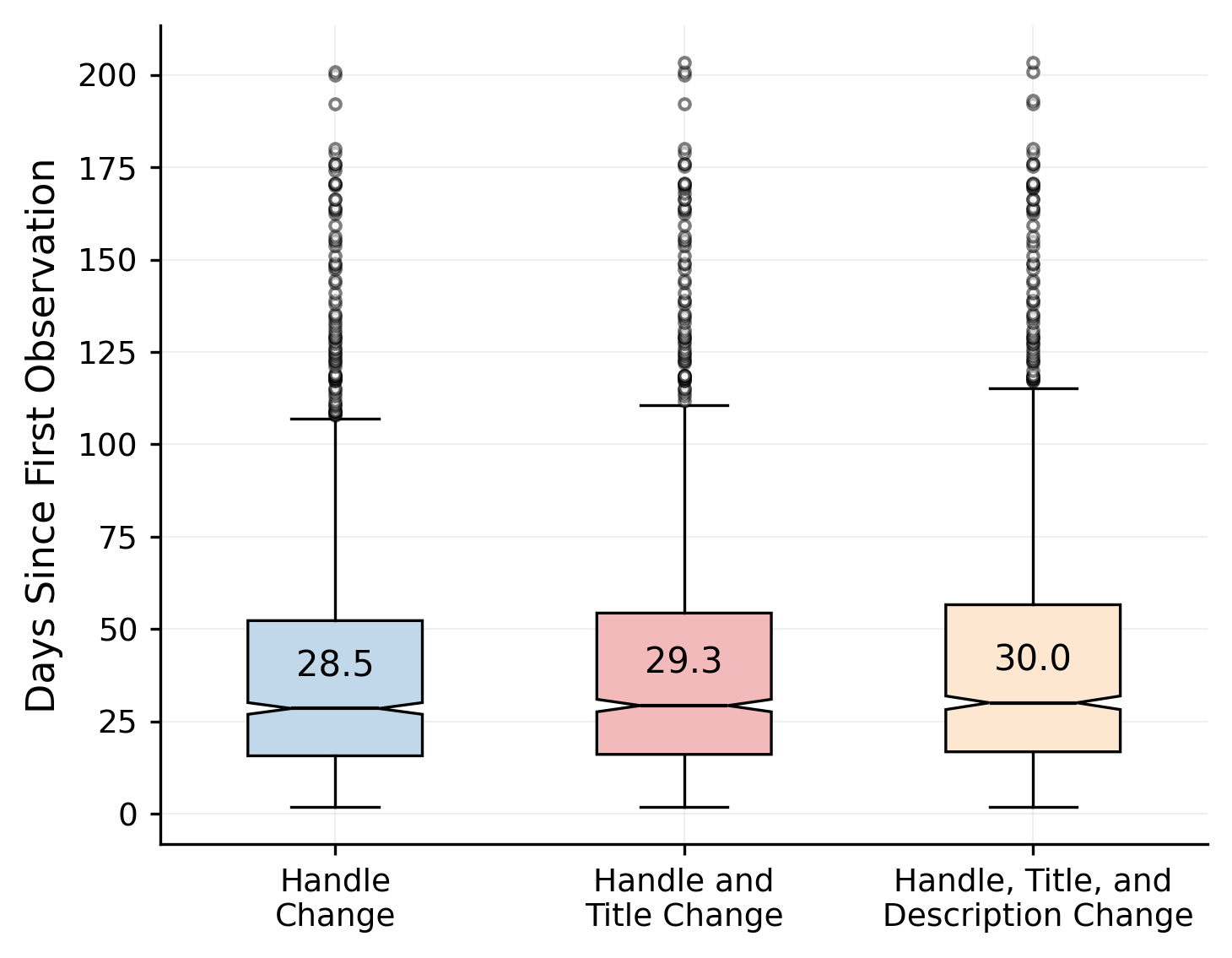}
    \caption{Days since a channel is listed for sale, until it changes its handle, title, and description (cumulative).}
    \label{fig:time_until_change}
\end{figure}
% Generated with /for_paper/time_until_change_distributions.py

\noindent\textbf{Time Until Repurpose.} For each listed channel, we computed the time it took
them to change their handle, title, and description. As seen in
Figure~\ref{fig:time_until_change}, the median time for a handle change
was 28.5 days, followed by 29.3 days for a change in the handle and
title, and 30 days for a change in handle, title, and description. Interestingly, handle, title, and description changes are not instantaneous, but instead occur over a period of 36 hours (1.5 days), a fact that could inform an anomaly detection system. These results can be interpreted as a proxy for the median time of a sale.

\begin{figure}[t]
    \centering
    \includegraphics[width=\linewidth]{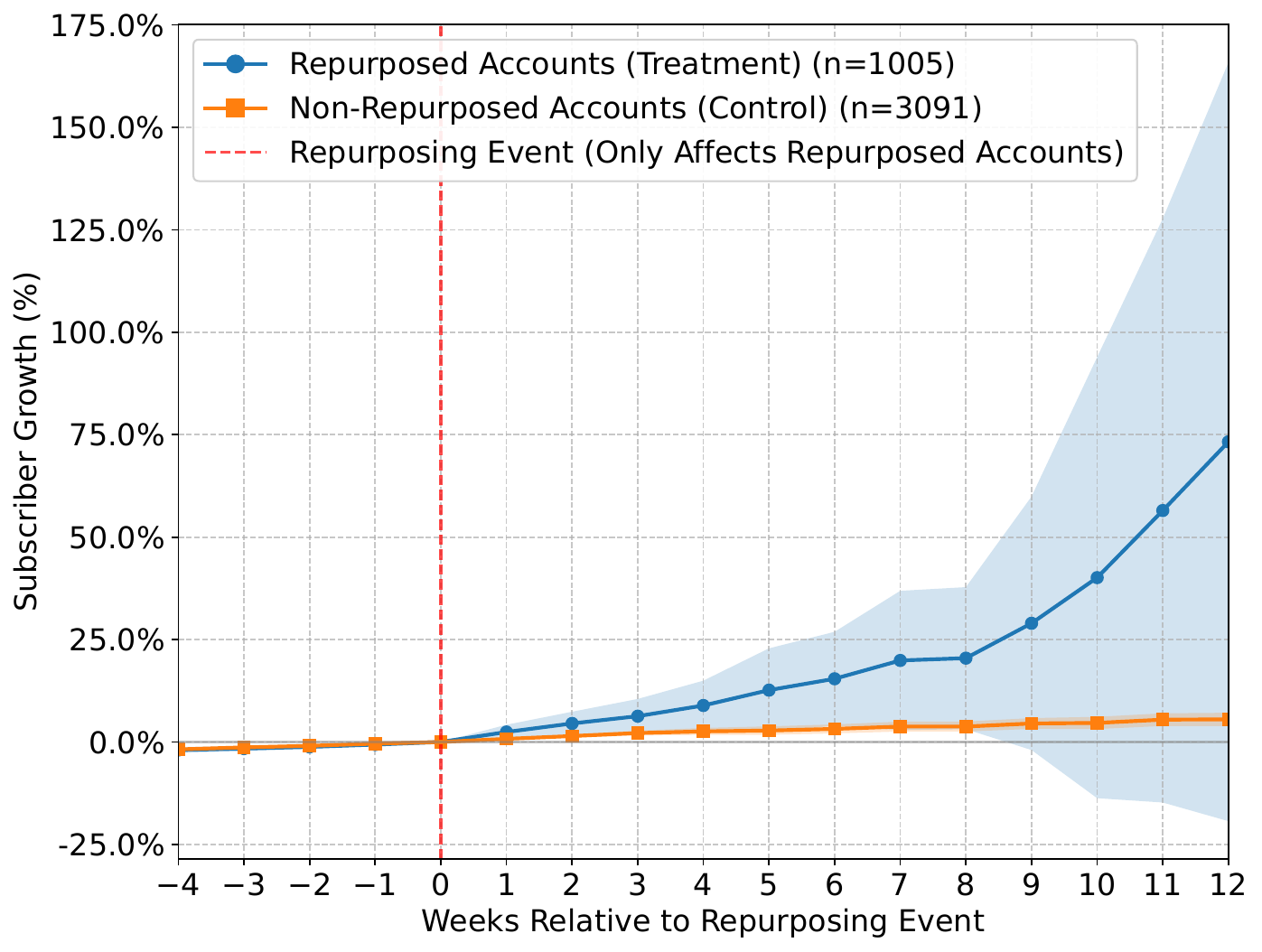}
    \caption{Subscriber growth comparison between repurposed (treatment) and non-repurposed channels (control). X-axis is relative to repurposing event. For non-repurposed accounts, the X-axis is based on the time they were first listed on Fameswap. Percentage growth is relative to the number of subscribers at $t$=0.}
    \label{fig:postchange}
\end{figure}
% Generated with /for_paper/subscribers_post_change.py

\noindent\textbf{Impact of Repurposing on Subscribers.} 
To assess the impact of repurposing on subscriber count, we aligned all time
series so that $t=0$ marks the week of the channel’s change
(Figure~\ref{fig:postchange}). We tracked subscriber change (\%) over the four
weeks before and 12 weeks after this point. We compared the subscriber growth
against a control group composed by channels from Fameswap that were not
repurposed. For these channels, we defined $t$=-4 as the first week when listed on Fameswap and we observed them for a total of 16 weeks (up to $t$=12).

Most channels show slight but steady growth prior to the change, likely due to
residual recommendation rather than active content updates. This pattern persists in the control group. However, repurposed channels'
subscriber counts increase on average--—often substantially. While some unsubscribes may be masked by new subscribers, the
consistent growth suggests most users remain unaware of the change and stay
subscribed. Lastly, on average we do not observe sharp changes in subscriber counts, which have been associated with the purchase of artificial engagement services (e.g., buying bot followers in bulk~\cite{nevado-catalan2023analysis}), or a large decrease due to a bulk suspension of bot accounts.

\begin{figure}[]
    \centering
    \includegraphics[width=\linewidth]{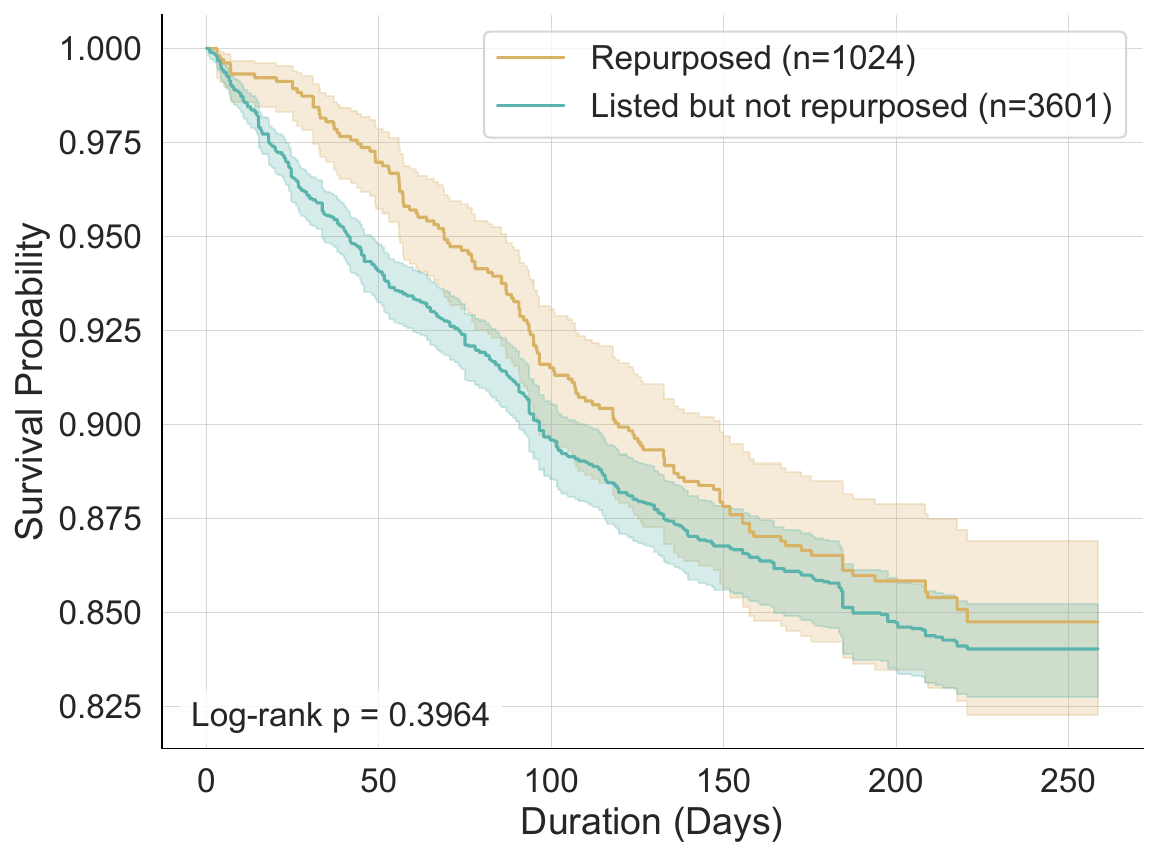}
    \caption{Survivability curve for repurposed and non-repurposed Fameswap channels. ``Death'' event is channel removal. Bands represent 95\% CI.}
    \label{fig:fswap_survivability}
\end{figure}

\noindent\textbf{Rate of Suspension.} We estimate the rate at which YouTube suspends Fameswap accounts, repurposed or otherwise. To do so, we compute the survival function for two groups, repurposed channels and those listed but not repurposed during our observation period, using the Kaplan-Meier estimator. As shown in Figure~\ref{fig:fswap_survivability}, all Fameswap channels exhibit an average survivability of approximately 85\% after 200 days (95\% CI: 79\%–87.5\%). Based on a log-rank
test, there is no significant difference between repurposed channels and those
that have not yet been repurposed.
Given that the unofficial sale of accounts likely violates YouTube's guidelines, we expected a higher rate of suspensions. However, these results indicate that repurposed channels are currently avoiding penalties.

\begin{figure}[t]
    \centering
    \includegraphics[width=\linewidth]{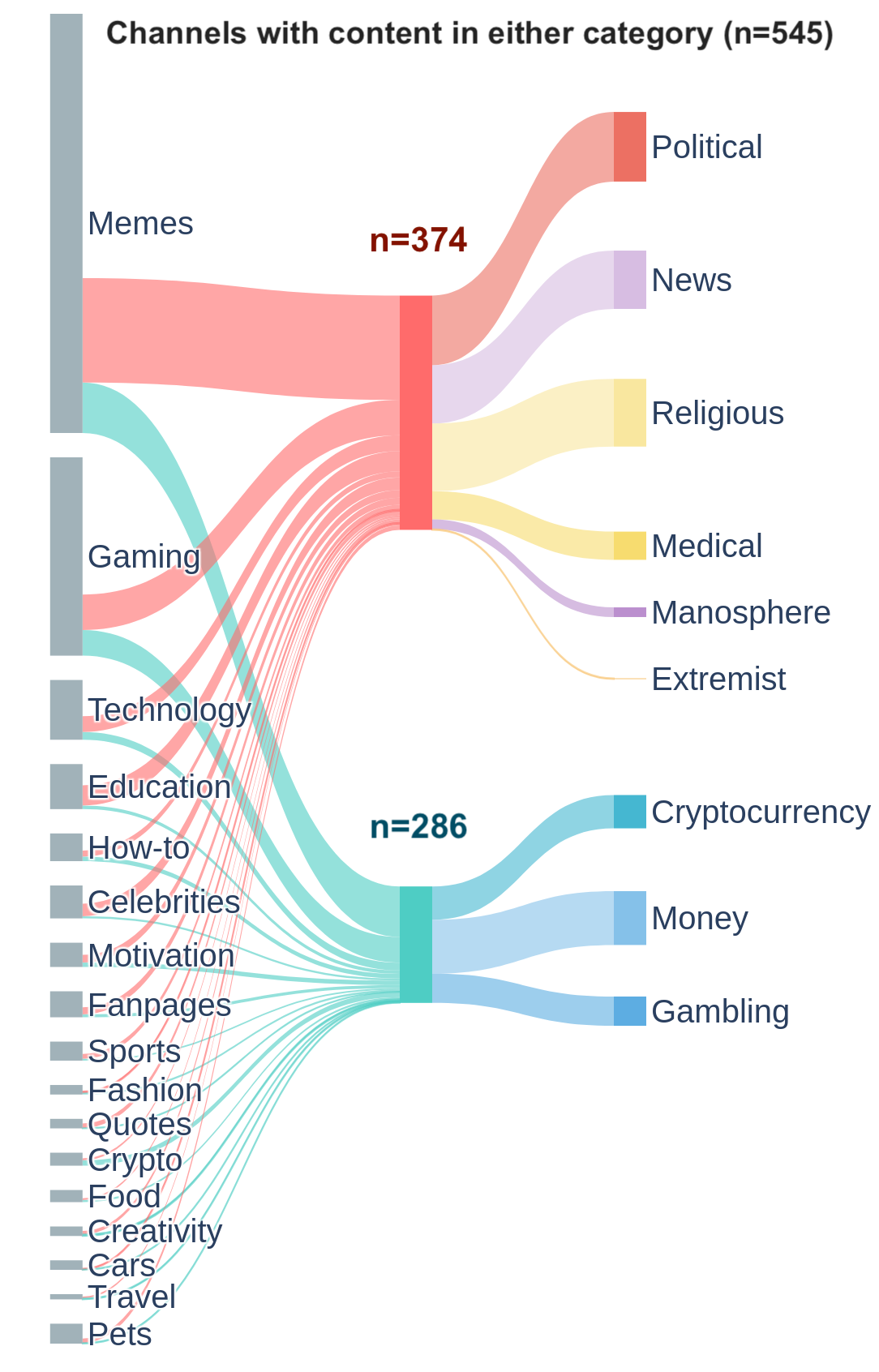}
    \caption{Topic presence for reused Fameswap channels. Topics on the left are self-reported at the time of listing. Topics on the right are labels assigned during our topic detection. Flows are proportional and one-to-many. That is, if a channel has several topics, flows are drawn to all destination topics proportionally. Counts overlap because some channels may have presence in both categories.}
    \label{fig:sankey}
\end{figure}
% Generated with /for_paper/sankey_diagram.py
\noindent\textbf{Changes in Content.}
We trace how channels transition from their original content category to policy-sensitive content after being repurposed. Using self-reported Fameswap categories (e.g., ``gaming'', ``technology'', etc.) and annotations after a channel has been repurposed, we highlight how content shifts.

When listing a channel for
sale on Fameswap, users choose a category that best fits the channel's
content. Choosing a category improves the searchability of the channel
in the marketplace. As observed in Figure~\ref{fig:fswap_distributions},
there are 18 categories that cover a wide range of topics. More than 88\% of the listings have categories. We manually verified the validity of these self-assigned categories and found them to be largely accurate. Using these self-assigned
categories, we investigated transitions in content categories, as shown in Figure~\ref{fig:sankey}. We find that
545 (53\%) of repurposed channels later displayed policy-sensitive content in their
channels. Of these, 374 had ideological content and 286 had financial
content (non-exclusive). In particular, the origin category (from Fameswap) had little to
do with the types of content that would appear in the
channel. We see that every category, even though seemingly benign (e.g., ``motivation'', ``food'') can shift to posting policy-sensitive content.

Not all channels began posting policy-sensitive content. During our observation period, there were many channels
in which we detected no policy-sensitive content even after being repurposed ($n=$479). For
example, a sports page became a musician's page, and an entertainment
page later became an influencer's travel vlog. Some
repurposed channels would later become channels associated with personal or commercial brands. Although not posting policy-sensitive content, these transitions raise a question about transparency across channels with large followings, given that people may rely on metrics such as subscriber counts to judge the authenticity of a brand.

\section{Generalizability and Prevalence}
\label{sec:gen-and-prev}

\begin{table*}[h!]
\centering
\caption{\label{tab:combined_descriptives} Regression results and descriptives for Fameswap and Socialblade models.}
\begin{threeparttable}
\setlength{\tabcolsep}{4pt}
\resizebox{\textwidth}{!}{%
\begin{tabular}{@{}lcccccccc@{}}
\cmidrule(lr){2-3} \cmidrule(l){4-9}
& \textbf{Fameswap} & \textbf{Socialblade} & \multicolumn{2}{c}{\textbf{Fameswap} (n=1,024)} & \multicolumn{2}{c}{\textbf{Socialblade} (n=1,047)} & \multicolumn{2}{c}{\textbf{Baseline} (n=2,972)} \\
& Coefficient (SE) & Coefficient (SE) & Mean & SD & Mean & SD & Mean & SD \\
\cmidrule(lr){1-3} \cmidrule(l){4-9}
(Intercept) & $\phantom{-}0.7093^{***}$ $(0.0994)$ & $\phantom{-}0.3401^{***}$ $(0.0824)$ & -- & -- & -- & -- & -- & -- \\
View Ct. (Mean) & $\phantom{-}0.0000\phantom{^{***}}$ $(0.0000)$ & $\phantom{-}0.0000\phantom{^{***}}$ $(0.0000)$ & 114{,}474 & 615{,}572 & 119{,}214 & 1{,}016{,}701 & 211{,}258 & 1{,}085{,}125 \\
View Ct. (SD) & $\phantom{-}0.0000\phantom{^{***}}$ $(0.0000)$ & $\phantom{-}0.0000\phantom{^{***}}$ $(0.0000)$ & 379{,}984 & 1{,}298{,}554 & 195{,}928 & 1{,}100{,}404 & 442{,}672 & 2{,}061{,}892 \\
Like Ct. (Mean) & $\phantom{-}0.0000^{***}$ $(0.0000)$ & $\phantom{-}0.0000\phantom{^{***}}$ $(0.0000)$ & 4{,}085 & 19{,}742 & 1{,}120 & 8{,}357 & 1{,}657 & 9{,}986 \\
Like Ct. (SD) & $\phantom{-}0.0000^{***}$ $(0.0000)$ & $\phantom{-}0.0000\phantom{^{***}}$ $(0.0000)$ & 12{,}994 & 53{,}193 & 1{,}761 & 9{,}201 & 3{,}290 & 15{,}435 \\
Comm. Ct. (Mean) & $-0.0007^{*}\phantom{**}$ $(0.0005)$ & $-0.0005\phantom{^{***}}$ $(0.0004)$ & 54.78 & 301.1 & 45.09 & 270.5 & 88.28 & 400.0 \\
Comm. Ct. (SD) & $\phantom{-}0.0001^{*}\phantom{**}$ $(0.0002)$ & $\phantom{-}0.0000\phantom{^{***}}$ $(0.0002)$ & 180.41 & 982.0 & 73.72 & 303.6 & 166.17 & 752.0 \\
Vid. Len. (Mean) & $\phantom{-}0.0001^{*}\phantom{**}$ $(0.0000)$ & $\phantom{-}0.0000\phantom{^{***}}$ $(0.0000)$ & 739.57 & 2{,}821.4 & 518.60 & 1{,}207.4 & 616.16 & 1{,}113.2 \\
Vid. Len. (SD) & $\phantom{-}0.0000^{***}$ $(0.0000)$ & $\phantom{-}0.0000\phantom{^{***}}$ $(0.0000)$ & 1{,}316.0 & 12{,}617.2 & 580.82 & 1{,}489.1 & 803.19 & 8{,}011.8 \\
T. Btw. Vids. (Mean) & $-0.0039^{*}\phantom{**}$ $(0.0022)$ & $\phantom{-}0.0002\phantom{^{***}}$ $(0.0002)$ & 17.59 & 58.93 & 119.73 & 328.23 & 72.11 & 213.58 \\
T. Btw. Vids. (SD) & $\phantom{-}0.0050^{***}$ $(0.0013)$ & $\phantom{-}0.0022^{***}$ $(0.0003)$ & 41.64 & 108.8 & 206.89 & 381.47 & 115.11 & 200.46 \\
Vids./Wk. (Mean) & $\phantom{-}0.0046\phantom{^{***}}$ $(0.0055)$ & $\phantom{-}0.0193^{*}\phantom{**}$ $(0.0117)$ & 5.65 & 9.03 & 3.01 & 5.59 & 3.65 & 18.43 \\
Vids./Wk. (SD) & $\phantom{-}0.0058\phantom{^{***}}$ $(0.0048)$ & $-0.0047\phantom{^{***}}$ $(0.0053)$ & 4.98 & 10.20 & 2.37 & 6.22 & 3.08 & 20.60 \\
T. 1st to Oldest Vid. & $\phantom{-}0.0289^{***}$ $(0.0050)$ & $\phantom{-}0.0224^{***}$ $(0.0027)$ & 17.78 & 23.15 & 57.72 & 64.77 & 84.15 & 62.78 \\
Oldest Vid. Age & $-0.0744^{***}$ $(0.0045)$ & $-0.0375^{***}$ $(0.0025)$ & 20.10 & 23.43 & 65.78 & 68.18 & 132.43 & 56.53 \\
Subs. Count & $\phantom{-}0.0000\phantom{^{***}}$ $(0.0000)$ & $\phantom{-}0.0000^{*}\phantom{**}$ $(0.0000)$ & 188{,}898 & 1{,}016{,}056 & 34{,}056 & 169{,}278 & 103{,}427 & 1{,}002{,}724 \\
Total Videos & $-0.0001^{*}\phantom{**}$ $(0.0001)$ & $-0.0002\phantom{^{***}}$ $(0.0001)$ & 162.76 & 493.97 & 127.09 & 806.19 & 373.66 & 1{,}365.7 \\
\cmidrule(lr){1-3} \cmidrule(l){4-9}
\textit{Content Features} & & & & & & & & \\
\cmidrule(lr){1-3} \cmidrule(l){4-9}
Non-YT Money & $\phantom{-}0.5092^{***}$ $(0.1464)$ & $\phantom{-}0.0713\phantom{^{***}}$ $(0.0936)$ & 0.347 & 0.476 & 0.372 & 0.483 & 0.432 & 0.495 \\
AI-generated & $\phantom{-}1.1557^{***}$ $(0.2087)$ & $\phantom{-}0.6283^{***}$ $(0.2000)$ & 0.112 & 0.316 & 0.066 & 0.248 & 0.022 & 0.147 \\
Political & $\phantom{-}0.1697\phantom{^{***}}$ $(0.1799)$ & $-0.1145\phantom{^{***}}$ $(0.1303)$ & 0.217 & 0.412 & 0.171 & 0.377 & 0.187 & 0.390 \\
Religious & $\phantom{-}0.9681^{***}$ $(0.1645)$ & $\phantom{-}0.3910^{***}$ $(0.1187)$ & 0.208 & 0.406 & 0.194 & 0.396 & 0.154 & 0.361 \\
News & $\phantom{-}0.7261^{***}$ $(0.2001)$ & $\phantom{-}0.4305^{***}$ $(0.1356)$ & 0.183 & 0.387 & 0.154 & 0.361 & 0.129 & 0.335 \\
Medical/Health & $-0.4594^{**}\phantom{*}$ $(0.1921)$ & $-0.3025^{*}\phantom{*}$ $(0.1300)$ & 0.096 & 0.294 & 0.121 & 0.327 & 0.153 & 0.360 \\
Cryptocurrency & $\phantom{-}1.7541^{***}$ $(0.2622)$ & $\phantom{-}0.5112^{*}\phantom{**}$ $(0.2424)$ & 0.116 & 0.321 & 0.039 & 0.194 & 0.014 & 0.118 \\
Gambling & $\phantom{-}1.4821^{***}$ $(0.3274)$ & $\phantom{-}0.2759\phantom{^{***}}$ $(0.2728)$ & 0.056 & 0.229 & 0.021 & 0.143 & 0.017 & 0.129 \\
Money/Stocks & $\phantom{-}0.2390\phantom{^{***}}$ $(0.1892)$ & $\phantom{-}0.1741\phantom{^{***}}$ $(0.1451)$ & 0.162 & 0.369 & 0.113 & 0.316 & 0.073 & 0.260 \\
Kids & $\phantom{-}0.6960^{**}$ $(0.2275)$ & $\phantom{-}0.3357^{**}\phantom{*}$ $(0.1341)$ & 0.099 & 0.298 & 0.1242 & 0.3299 & 0.075 & 0.264 \\
P. Copyright Infr. & $\phantom{-}0.5022^{***}$ $(0.1517)$ & $\phantom{-}0.2826^{**}\phantom{*}$ $(0.0943)$ & 0.254 & 0.436 & 0.345 & 0.476 & 0.426 & 0.495 \\
Manosphere & $\phantom{-}0.8533^{*}\phantom{**}$ $(0.4041)$ & $\phantom{-}0.4109\phantom{^{***}}$ $(0.4444)$ & 0.036 & 0.187 & 0.011 & 0.106 & 0.006 & 0.078 \\
\cmidrule(lr){1-3} \cmidrule(l){4-9}
\end{tabular}
}%
\begin{tablenotes}
\footnotesize
\item Notes: $^{***}p<0.001$, $^{**}p<0.01$, $^{*}p<0.05$ ; Overlapping channels across samples were excluded.
\end{tablenotes}
\end{threeparttable}
\end{table*}

We investigate the prevalence of channel repurposing and whether channel repurposing is observable beyond the accounts that are traded in social media marketplaces like Fameswap. By using a random sample of channels from the analytics platform Social Blade, we estimate the prevalence of repurposing ``in-the-wild.'' Furthermore, using a logistic regression model we investigate which channel features correlate with repurposing. We confirm that repurposed accounts, beyond those that are commercialized in open markets, are also used to disseminate the content we identified in \S~\ref{sec:content_analysis}. In addition, we find that repurposed channels tend to have more subscribers on average. Lastly, posting behavior (e.g., upload cadence) and content are most predictive of repurposing. We summarize the descriptive statistics of our three samples (repurposed Fameswap channels, repurposed Social Blade channels, and a random sample of non-repurposed Social Blade channels) in
Table~\ref{tab:combined_descriptives}.

\subsection{Prevalence of Repurposed Channels in the Wild}
\label{subsec:prevalence}

We labeled the content of the large sample of YouTube channels we collected in \S~\ref{subsubsec:socialblade_channels}, using the annotation procedure in \S~\ref{sec:channel_reuse}. Of the 1.4M YouTube channels, we detected 10,200
(0.73\%) channels that changed their handle between January and
March 2025. Of these, our LLM labeled 3,584 channels as repurposed. To validate the annotations, we randomly
sampled 10\% of the channels ($n$=1,020) and manually annotated each
pair of observations according to our definition. The false positive and negative rates are 4.9\% and 3.8\%, respectively.

The agreement between our annotations and the LLM classifier is 95.54\%. Our classifier had an accuracy between 94.2\% and 96.6\% (95\% CI), resulting in an estimated 3,384–3,456 repurposed accounts.
These results indicate that, from Jan. 2025 to Mar. 2025,
approximately 0.24-0.25\% of channels in the Social Blade population
(68M+ channels) were repurposed. Of the repurposed channels we observed, 1,074 had
more than 1,000 followers. These repurposed channels had a
total audience of 43,975,420 subscribers, a sizable audience. These results imply that
there were 160,000+ repurposed channels ($\sim$0.24\% of the 68M captured
by Social Blade) that we did not capture, likely with tens or hundreds
of millions of subscribers.

\subsection{Indicators of Repurposing}
\label{sec:indicators_of_reuse}
We use a logistic regression model to examine the relationship between
channel characteristics, channel content, and channel repurposing. We model the log-odds
of the outcome as a linear combination of the predictors:
\begin{multline}
\log\left(\frac{p}{1-p}\right) = \beta_0 + \beta_1F_1 + \cdots + \beta_jF_j \\
+ \beta_{j+1}T_1 + \cdots + \beta_{j+k}T_k \ , \nonumber
\end{multline}
where $\beta_0$ is the intercept (log-odds outcome when all predictors
are zero), $F_i$ represents quantitative features extracted from each
channel, and $T_i$ represents an indicator for each content category. In this
model, a one-unit change in a covariate $X_i$ changes the log-odds
outcome by $\beta_i$; equivalently, the odds change by a factor of
$e^{\beta_{i}}$.

\noindent\textbf{Quantitative Features.}
We take the mean and standard deviation of all observed video views,
likes, and comments. We also compute the mean and standard deviation
of the number of videos posted weekly, video lengths (seconds), and
the time between each video posted (days). Finally, we compute the time
between the first and last video the channel has posted (months) and the
oldest video the channel has, which would typically mark when the channel first started uploading content.
Finally, we take the total number of videos. Table~\ref{tab:combined_descriptives} presents the descriptive statistics
for each sample.
Video deletions are likely an important feature. However,
because we do not have frequent snapshots of Social Blade channels
(given the sample size), we miss all uploads and deletions that take
place between the two snapshots---a limitation that the time between videos may still partially address.

\noindent\textbf{Qualitative Features.} For content, we encode topic presence as indicator variables derived from our LLM categorization and gold-standard labels (described in \S~\ref{sec:content_analysis}).

\subsection{Regression Results}

We conduct two logistic regressions, first we model repurposed channels
from Fameswap against a baseline drawn from Social Blade using the \texttt{R} package, ~\texttt{DSL}~\cite{dslrpackage}. We report the
coefficients in Table~\ref{tab:combined_descriptives}.

\noindent\textbf{Takeaways from Quantitative Features:} 
Channels whose oldest video is more recent
are more likely to be repurposed, as are those with a longer interval between the oldest and most recent uploads.
So, buyers seemingly favor channels that look established
(i.e., showing a sizeable upload time span), but those are not necessarily truly long-standing: a genuinely mature
channel would feature \emph{both} an old first upload and a long production span.

Irregular upload intervals (high SD in
``time-between-videos'')
predict repurposing, implying that erratic schedules and an existing
audience are typical markers of repurposed channels. Average views across videos were not significantly different across samples. Repurposed Fameswap channels tended to have a higher number of likes in videos, a greater variance in likes in videos, and lower
comment counts. It is possible that repurposed channels may disable or limit
comments to avoid scrutiny. The likes and views pattern did not hold for repurposed channels found in the wild. However, in both cases subscriber counts were not a
predictor of channel reuse, indicating that channels of all sizes are repurposed. However, sold channels tended to have higher subscriber counts possibly because they are more appealing for sale.

\noindent\textbf{Takeaways from Qualitative Features:} Repurpose odds
are primarily content-driven. AI-generated content ($\beta$=0.6–1.1) is
the strongest predictor of repurposing, probably because automated
pipelines scale cheaply after a channel is acquired. However, we rely on AI usage disclosure
to tag channels, as described in \S~\ref{sec:content_analysis},
so the coefficient may also reflect self-disclosure bias.
Cryptocurrency shows the next-largest effect ($\beta$=1.7 on Fameswap),
followed by gambling. We hypothesize that these categories are
prone to reuse because of their monetization potential. In addition, dubious projects may repurpose accounts to be perceived as legitimate. In line
with previous work, we find a significant amount of off-platform
monetization~\cite{hua2022characterizing,chu2022tube}, albeit only
for Fameswap. Medical/health is negatively associated,
suggesting either stricter platform scrutiny or lack of prominent
health-related narratives. Religious
content and potentially copyright-infringing content are associated
with repurposing. A possible explanation could be that these types of
content are used to grow channels due to their appeal (e.g., free TV
shows), similar to kids content, and we are identifying trace videos
prior to deletion.

\section{Discussion and Conclusions}
Our results suggest that channel repurposing involves sold channels
(such as those advertised on Fameswap), as well as channels that may
not have been for sale (such as most of those found through Social Blade).
Across most Fameswap accounts, channel repurposing 
seemed to go unnoticed both by users and YouTube, as indicated by the subsequent
growth in subscribers and lack of suspension. The median time between a channel being listed for sale and being repurposed was 30 days.
However,
some took as long as 200+ days (\textbf{RQ1}).

In both
samples, we find that channels with large numbers of subscribers---193M+
for Fameswap and 43M+ for Social Blade---completely repurposed their
channels, erasing any perceivable association with their prior identity.
By leveraging a large sample of YouTube channels, from a random 
Social Blade sample, we estimated that 0.24-0.25\% of channels were
repurposed between January and March 2025. These results imply that
there were another 160,000+ repurposed channels (out of the 68M captured
by Social Blade) that we did not capture, likely with tens or hundreds
of millions of subscribers  (\textbf{RQ2}).

Engagement metrics offer little indication that a
channel has been or will be repurposed. Instead, video upload behaviors
(i.e., time between videos) and potential gaps in their video history
are better indicators for potential repurposing. These indicators, unfortunately, may not be detectable or readily interpretable by everyday
users. Furthermore, usage of AI was also a significant predictor of channel repurposing, indicating that ``faceless channels'' are commonly repurposed (\textbf{RQ3}).

The presence of policy-sensitive content in the channel was also a significant predictor in both samples.
On Fameswap, 53\% of channels posted potentially deceptive
ideological and financial content, as per YouTube's guidelines. Before repurposing, they belonged to various innocuous content categories,
ranging from humorous content to sports and celebrity gossip. 
Figure~\ref{fig:2_part_real_example} illustrates
an
example of a problematic transition, in which a channel that previously
shared ``interesting facts'' is sold and repurposed into a news channel
featuring contentious geopolitical actors: Russia, Ukraine, Zelenskyy,
etc. Due to space constraints, we cannot discuss all remarkable cases
individually, and instead, quantitatively show that policy-sensitive content is significantly related to channel repurposing
(\textbf{H1}).

Our results suggest that social media audiences are becoming more
commoditized (i.e., have less differentiation, have widespread
availability, and compete mainly on price). From the lens of transaction
cost economics, social media account markets are evolving from forums
(unassisted markets) to assisted markets, with lower barriers to
entry and increased liquidity~\cite{williamsonTCE}. This transition
typically makes outsourcing more attractive and, as a result, fosters
specialization, a pattern that frequently emerges in cybercriminal
endeavors~\cite{wegberg2018plug,stringhini2014harvester,huang2015framing}. In our context, this may signal the emergence of actors who
specialize in harvesting organic audiences. 

We do not claim nor expect
these markets to be free from fraud. Many vendors may use artificial engagement to inflate their numbers and sell accounts
at a premium. However, these markets are growing. They are also increasingly
becoming more sophisticated, allowing sellers to share screenshots and incorporating feedback systems. This continuous evolution, coupled with the apparent success
of these markets, seems to indicate that a substantial portion of buyers
find these markets useful.

We found evidence that repurposed channels may play a role in influence operations, which, similar to past work, is driven by content features~\cite{alizadeh2020content}. We confirm this finding in the context of YouTube by uncovering various channels that turned into journalistic-like channels. Additionally, generative artificial intelligence (GenAI) 
seems to play an important role in the automation of content
creation to harvest audiences, in line with recent small-scale investigations~\cite{diRestaGoldstein2024}.
Putting misinformation consumption in terms of supply and demand,
scholars have argued that AI does not increase the demand for
misinformation~\cite{simon2023misinformation}. However, our results add
nuance to these claims. By facilitating the cultivation of audiences
that can then be served mis/disinformation, GenAI may be helping
increase the demand for disinformation indirectly. Furthermore,
GenAI may enable actors to more easily produce diverse
content across niches and languages, facilitating the cultivation of
specific audience demographics---a strategy that can be employed to target specific populations when selecting channels for repurposing.

There are a few possible interventions. First, our results highlight a set of features that distinguish repurposed channels from benign ones. Second, we recommend increased transparency.
There are many legitimate reasons to change channel handles and titles, and doing so without drawing attention is somtimes advisable.
However, we show that current imperceptibility can be exploited to the detriment of users. Thus, if a channel is large enough, we suggest that subscribers be notified of meaningful changes in identity and content, and that historical handle changes and titles be auditable. Another option is to 
monitor second-hand social media account marketplaces and seek to
deplatform them. We do not suggest this approach. These interventions often have short-lived benefits and make nefarious activity harder 
to measure~\cite{ahnvu2024deplatforming}. Lastly, another option is to make monetization harder, a strategy that YouTube adopted on July 15th, 2025, to target inauthentic content~\cite{YouTubeMonetizationPolicies2025}. However, many accounts already leverage various forms of off-platform monetization. Out of these interventions, we are most optimistic about approaches that increase transparency and user agency.

\subsection{Limitations and Future Work}
\label{sec:limitations}
We rely on video metadata to
infer topics, without analyzing the video content itself. Thus, when
labeling channels as political, medical, or news-related, we do not
assess whether the content is actually harmful or mis/disinformation. Automated fact-checking is a hard and open problem~\cite{xiao2025sok}, particularly for developing events. In addition,
fact-checking video content remains underexplored, with most of the work
focused on news articles~\cite{guo2022survey}. Nonetheless, the veracity of the claims does not preclude the usefulness of these channels as vectors of influence.

We only checked for account suspension as a punitive measure. However, de-monetization likely precedes suspension. To the best of our knowledge, there is currently no reliable method to check monetization status, which inhibited us from verifying whether YouTube is currently detecting repurposed channels and de-monetizing them. However, even if this is the case, account marketplaces have only gotten more popular, which puts into question whether de-monetization is a sufficient deterrent.

Assessing harm is
subjective and verifying whether exposure to content led to harm is often impossible. While not all cryptocurrency or money-themed
channels promoted obvious scams, past work has financial loss back to influencers' advice~\cite{kawai2023is}. 
To assess claims in videos,
future work could analyze transcripts, video frames, or external URLs
to refine these labels and estimate the presence, volume, and degree of harmful content.

Fameswap seems to be among the leading marketplaces~\cite{beluri2024exploration}.
However, the extent of overlap across marketplaces or between
marketplaces and forums is unknown. This motivated our prevalence
estimation, which avoids limiting analysis to marketplace samples.
Finally, our study focuses only on YouTube. While we
expect our findings to generalize to platforms like TikTok, Twitch, or
Twitter/X, this remains to be tested.

\cleardoublepage
\noindent\textbf{Acknowledgments.} This research was partially supported by the Singapore Defence Science and Technology Agency (DSTA) under agreement CNZ2000832. It benefited from great work by Haoxiang Yu. Lastly, immense thanks to our external social media expert, Asia Grant.
\appendix

\section{Ethical Considerations}
\label{sec:ethics}

This study involved various stakeholders and careful ethical consideration of various decisions, particularly data collection. Ultimately, our decision to conduct and publish this study is guided by the fact that it is the first to document repurposing of social media accounts, which we find may affect millions of people. All data collected were stored on a private server hosted at our university and accessed only by members of the research team, unless otherwise indicated. 

\noindent\textbf{Fameswap.} Fameswap is a website allegedly registered in the US with 12 years of activity. Purchasing a Fameswap membership raises ethical concerns, as it supports a service that some individuals can use to carry out policy-violating practices, as we found in this study. However, Fameswap is \textit{not illegal} nor do we claim that Fameswap is an active collaborator in the practices we uncovered. However, as long as 1) alternative options are not available, and 2) the monetary amounts considered are small---especially when compared to the potential scientific value of developing countermeasures---there is precedent in the research community, even for goods and services that are illegal, such as ``booters'' (DDoS-for-hire services)~\cite{Karami:WWW2016}, goods from spam-advertised websites~\cite{Levchenko:S&P11,Kanich:Sec11}, and even hacking services~\cite{mirian2019hack}, with purchases surpassing a thousand dollars. These numbers are in line with or exceed our own expenses: USD~348 for a monthly and yearly subscription.

The Fameswap platform does not explicitly prohibit scraping and there is also recent precedent of account registration to access data, even in problematic platforms~\cite{campobasso-know,han2025characterizing,gibson2024analyzing}. The site's \texttt{robots.txt} disallows nothing (i.e., \texttt{Disallow: <empty>}). Our web scraping activities added to server load and bandwidth usage; however, we collected at most tens of pages per day; we believe our impact was negligible.

We considered whether to publish Fameswap's name, as it may attract attention to it. We decided to name Fameswap as it has been named in recent work~\cite{chu2022tube,beluri2024exploration} and may encourage researchers to study similar platforms and other commercialized accounts beyond YouTube. Distribution of the data we collected from Fameswap, however, merits scrutiny. We expand on this point in \S~\ref{sec:open_science}.

\noindent\textbf{Users of the Fameswap Marketplace.} In this study, we collected metadata from YouTube accounts listed for sale on Fameswap, as well as metadata from those accounts through YouTube. We did not attempt to deanonymize any Fameswap user nor did we employ user-level data beyond counting the number of vendors on the platform. We did not interact with users (e.g., via direct messages). We do not report quotes. We blur account and channel identifiers, where necessary, to prevent driving traffic to these channels and for user privacy.

\noindent\textbf{Social Blade and the Internet Archive.} Similar to Fameswap, we sought to mitigate our impact on the platforms from
which we collected data: the Wayback Machine and Social Blade. Where possible, we opted to use existing data, rather than query it
ourselves. We did not scrape Social Blade, but instead sampled
from the data that Horta Ribeiro and West had
collected~\cite{hortaribeiro2021youniverse}. To collect data from the Internet Archive, we followed their API guidance and limits~\cite{internetarchive_developers}. 

\noindent\textbf{YouTube.} We employed yt-dlp to assist in obtaining data from YouTube during this period. It is important to note that historically there has been some concern over the use of yt-dlp. In 2020, the
Recording Industry Association of America, issued a takedown notice to GitHub
under the Digital Millenium Copyright Act (DMCA), requesting the removal of the
project and its forks, arguing that it violated German copyright
law~\cite{youtube_dl_wikipedia}. Nonetheless, the takedown was
reversed~\cite{youtube_dl_wikipedia}, and is worth noting that the main concern was using yt-dlp for video downloads, not metadata (which was our use). To date, yt-dlp remains a useful tool for
academic projects, even for video downloads~\cite{retkowski-waibel-2024-text}. In our case, we only used yt-dlp to collect metadata from channels and chose conservative limits.
We originally intended to collect data through the YouTube Research program. We submitted a proposal prior to starting the study and received approval. Unfortunately, there was an issue with the API key that was provided and this was only resolved after the study period ended. %Thus, the data we collected as part of this study is not subject to YouTube Research program's policies, which has implications for data access (discussed in \S~\ref{sec:open_science}).

\noindent\textbf{YouTube Users.} For all YouTube channel data, we sought to incorporate the privacy guidelines and considerations
proposed by Beadle et al.~\cite{beadle2025sok_privacy}. Although YouTube does not meet Beadle et al.'s definition of social media data\cite{beadle2025sok_privacy, wikipedia_social_networking_services}, we still applied the same privacy considerations as with other social media sites. We did not use emails or phone numbers in our analysis, even though some accounts volunteer this information in their channel metadata. We also focused our analyses on larger channels ($>$1,000 subscribers), as we believe that they have a
lower expectation of privacy, given their large audience.

\noindent\textbf{Research Team.}
Labeling the content in each channel required data annotation which was conducted primarily by two members of the research team and one external member. All annotators were briefed about the potential content they would encounter. Each annotator participated voluntarily, eagerly, at their leisure, and with the freedom to stop at any point without repercussions. Because we only focused on metadata, the chances of encountering disturbing content were minimized.

\cleardoublepage
\section{Open Science}
\label{sec:open_science}

To conduct this study, we pre-registered our regression hypotheses, methods, and analysis plan prior to conducting the study. The pre-registration is available at \url{https://doi.org/10.17605/OSF.IO/FG9PM}. Our Zenodo repository with supplemental material is available at \url{https://doi.org/10.5281/zenodo.18216091}.

\subsection{Samples Considered in this Study}
\label{subsubsec:samples_considered}
The main samples that we considered in this study are:
% For submission
\begin{itemize}[leftmargin=*,itemsep=2pt,topsep=2pt]
    \item \textbf{Set of channels for sale on Fameswap}: All channels collected from Fameswap, $n$=4,641 from October 21 2024, to March 31 2025 (see \S~\ref{subsubsec:fswap_channels}).
    \item \textbf{Set of repurposed channels from Fameswap}: Subset of repurposed Fameswap channels, $n$=1,024 (see \S~\ref{sec:channel_reuse}).
    \item \textbf{Large random sample from Social Blade}: All Social Blade channels, appearing both in January and March 2025, $n$=1,351,912 (see \S~\ref{subsubsec:socialblade_channels}).
    \item \textbf{Set of repurposed channels from Social Blade ($>$1,000 subscribers}: Subset of Social Blade channels that we classified as repurposed, $n$=1,074 (see \S~\ref{subsec:prevalence}).
    \item \textbf{Random subsample from Social Blade (Baseline)}: Random subset of channels drawn from Social Blade with more than $>$1,000 subscribers, $n$=3000.
\end{itemize}

Out of these samples, the main ones needed to replicate the main findings of these experiments are the labeled and categorized repurposed channels from Fameswap ($>$1,000 subscribers, $n$=1,024), labeled and categorized repurposed channels from Social Blade ($>$1,000 subscribers, $n$=1,047), and the random sample from Social Blade ($>1,000$ subscribers, $n$=2,972). For the regression, we removed overlapping channels between samples, which reduced the final counts. We provide these dataframes along with scripts and documentation to replicate our main findings: Figure~\ref{fig:postchange}, Figure~\ref{fig:sankey}, Figure~\ref{fig:fswap_survivability}, and Table~\ref{tab:combined_descriptives}. These data also contain the handle changes per channel. These materials are found in the Zenodo repository.

\subsection{Supplemental Materials}
\label{apx:sup_materials}
To ensure that human coders
reviewed YouTube channels in a consistent way, with access to the same
information, we generated a markdown document for each channel, which included
the channel's handle changes, a timeline of the channel's profile information
over time, and all videos observed for the channel. The coders were instructed to read through this
document and use the coding guide (Appendix~\ref{apx:coding_guide}) to categorize the channel. We also used LLM prompts to annotate channel repurposing and topic categorization. Note, for the topic categorization prompt, we
used a few-shot prompting strategy, where we provided the LLM with examples of
each topic to help it understand the classification task. Due to space constraints, we provide pointers to these materials in the Zenodo repository.

\subsection{Data Requests}
The distribution of data we collected is governed by two main policies: YouTube's Data API policy~\cite{youtube_api_services_tos} and Fameswap's policy~\cite{fameswap_terms}, both of which may inhibit public data release. Interested parties may submit a request by contacting the authors. Requests will be reviewed based on several criteria: verifying the identity of the requester, corroborating that the usage is for academic purposes, confirming that the data sought is relevant to the stated purpose, and, when necessary, consulting external experts or our institutions if the request cannot be clearly resolved internally. If access is granted, we may impose conditions on its use.

\cleardoublepage
\small{
\bibliographystyle{plain}
\bibliography{99_bib}
}
\section{Appendix}

\subsection{Repurposed Channels Before Observation}
\label{apx:wayback}
To obtain a historical view of changes, we leveraged snapshots from the Wayback Machine~\cite{wayback_archive}. The Wayback Machine is a digital archive initiative by the Internet Archive; it allows users to go ``back in time'' to see how websites looked in the past. The Wayback Machine allows users to capture pages for archival purposes~\cite{wayback_wikipedia}. The frequency of snapshots varies per website.  More popular websites (e.g., higher-ranked by Alexa, higher number of inbound links, etc.) will be crawled more often~\cite{wayback_wikipedia}. Given a sample of YouTube channels, we attempt to obtain monthly snapshots from October 2022 to March 2025.

We queried the archive for snapshots taken after YouTube introduced handles (October 2022) for all Fameswap-listed channels (n=4,461), including those not yet repurposed, and all repurposed Social Blade channels (n=1,074).

For Fameswap, we found snapshots for 819 channels (17.6\%). Of these, 332 (40.5\%) had a different handle than the one listed at the time of sale, and 71 (9.5\%) showed more than one handle change across archived versions. Among the repurposed Social Blade channels, 289 (27.8\%) had snapshots; 116 (40.1\%) had previously used different handles, and 29 (10.0\%) had changed more than once. Due to the absence of additional metadata (e.g., channel descriptions), we could not assess whether these changes were substantial or cosmetic.

The frequency of handle changes may be meaningful. Prior work by Mariconti et al.~\cite{mariconti2017whats} found that Twitter accounts with multiple profile name changes were more likely to engage in misbehavior---a pattern that may hold for repurposed YouTube channels as well.

\subsection{Coding Guide}
\label{apx:coding_guide}

The coding guide we used to label content is shown in Table~\ref{tab:codebook}.

\begin{table*}[]
\caption{Coding guide for channel categorization.}
\begin{tabular}{@{}p{0.25\linewidth}p{0.63\linewidth}p{0.12\linewidth}@{}}
\toprule\toprule
\textbf{Did the channel contain:} & \textbf{The channel discusses or contains videos:} & \textbf{Motivation} \\ \midrule
Politically-related content? & Describing contemporary political subjects, events, or figures. & \cite{youtube2025misinformation,youtube2025elections} \\
News-related content? & Describing news reports and/or world events emulating traditional newscast. &~\cite{youtube2025misinformation,youtube2025elections}   \\
Health-related content? & Describing medical or health-related subjects, events, or figures. & \cite{youtube2025medical} \\
Religious content? & Describing religious subjects, events, or figures. & \cite{albadi2022deradicalizing,albadi2018are}\\
Manosphere content? & Discussing manosphere or redpill topics, events, or figures. & \cite{ribeiro2020auditing}\\
Extremist content? & Showcasing toxic content or content from extremist groups, such as white supremacist, jihadist, neo-nazi, alt-right, etc. & \cite{youtube2025violent_extremist}\\
Money-making content? & Describing investing and/or financial subjects, events, or figures, including trading, digital marketing, and e-commerce. & \cite{youtube2024spam} \\
Cryptocurrency-related content? & Describing cryptocurrency subjects, events, or figures. & \cite{youtube2024spam}\\
Gambling-related content? & Describing gambling and/or betting subjects, events, or figures, including sports gambling and online casinos. & \cite{youtube2024illegal}\\
AI-generated content? & Showcasing content created with generative AI tools, e.g., Midjourney, Sora, Runway, etc. & \cite{youtube_ai_2025}\\
Content that may infringe copyright? & Describing software, shows, movies, or music that they do not own, as well as advertising links or websites that contain these materials. & \cite{youtube2025copyright}\\
Content geared towards children? & Oriented towards very young audiences. For example, they contain videos about nursery rhymes and kids shows. & \cite{youtube2024child}\\
Alternative forms of monetization? & Containing information on how to purchase products/services from an external site, business inquiry emails and phone numbers, links to WhatsApp or Telegram groups advertised to make money. & \cite{chu2022tube,hua2022characterizing}\\
\bottomrule
\end{tabular}
\label{tab:codebook}
\end{table*}

\tcbset{
  annotationprompt/.style={
    colback=gray!5,
    colframe=black,
    arc=3mm,
    boxrule=0.4pt,
    fonttitle=\bfseries\small,
    title={Prompt: Channel Repurpose Annotation}
  }
}
\clearpage

%%%%%%%%%%%%%%%%%%%%%%%%%%%%%%%%%%%%%%%%%%%%%%%%%%%%%%%%%%%%%%%%%%%%%%%%%%%%%%%%
\end{document}